\documentclass[aps,pre,reprint,numerical,superscriptaddress,twocolumn,showpacs,amsmath,amssymb]{revtex4-1}

\usepackage{graphicx}
\usepackage{dcolumn}
\usepackage{bm}
\usepackage{color}
\usepackage{ulem}

\usepackage[latin1]{inputenc}
\usepackage{amsfonts}
\usepackage{amsmath}
\usepackage{amssymb}
\usepackage{amsthm}
\usepackage{bbold}
\usepackage[american]{babel}
\usepackage[T1]{fontenc}
\usepackage{xspace}
\usepackage{bbm}
\usepackage[all]{xy}
\providecommand{\begeq}[1]{\begin{equation}#1\end{equation}}

\begin{document}

\title{Self-consistent field theory based molecular dynamics with linear system-size scaling}
\author{Dorothee Richters}
\affiliation{%
Institute of Mathematics and 
Center for Computational Sciences, Johannes Gutenberg University Mainz, Staudinger Weg 9, D-55128 Mainz, Germany
}%

\author{Thomas D. K\"uhne}%
\email{kuehne@uni-mainz.de}
\affiliation{%
Institute of Physical Chemistry and 
Center for Computational Sciences, Johannes Gutenberg University Mainz, Staudinger Weg 7, D-55128 Mainz, Germany
}%

\date{\today}

\begin{abstract}
We present an improved field-theoretic approach to the grand-canonical potential suitable for linear scaling molecular dynamics simulations using forces from self-consistent electronic structure calculations. It is based on an exact decomposition of the grand canonical potential for independent fermions and does neither rely on the ability to localize the orbitals nor that the Hamilton operator is well-conditioned. Hence, this scheme enables highly accurate all-electron linear scaling calculations even for metallic systems. The inherent energy drift of Born-Oppenheimer molecular dynamics simulations, arising from an incomplete  convergence of the self-consistent field cycle, is circumvented by means of a properly modified Langevin equation.
The predictive power of the present linear scaling \textit{ab-initio} molecular dynamics approach is illustrated using the example of liquid methane under extreme conditions. 
\end{abstract}

\pacs{31.15.aq, 31.15.E-, 71.15.Dx, 71.15.Pd}
\keywords{ab-initio, Linear Scaling, Tight-Binding, Car-Parrinello, Molecular Dynamics, Methane}
\maketitle

\section{\label{sec:level1}Introduction}

\textit{Ab-initio} molecular dynamics (AIMD), where the forces are calculated on-the-fly by accurate electronic structure methods, has been very successful in explaining and predicting a large variety of physical phenomena and guiding experimental work \cite{marx2009}. However, the increased accuracy and predictive power of AIMD simulations comes at a significant computational cost, which has limited the attainable length and time scales in spite of recent progress \cite{kueh1, hutter}. 
As a consequence, Hartree-Fock (HF), density functional theory (DFT) \cite{kohn1} and even the semi-empirical tight-binding (TB) approach \cite{slater, seifert} are to date the most commonly used electronic structure methods in conjunction with AIMD. However, for large systems the calculation of the electronic structure and hence total energies as well as nuclear forces of atoms and molecules is still computationally fairly expensive. This is due to the fact that solving the Schr\"odinger equation is a high-dimensional eigenvalue problem, whose solution requires diagonalizing the Hamiltonian of the corresponding system, which typically scales cubically with its size. 
Therefore, a method that scales linearly with the size of the system would be very desirable, thus making a new class of systems accessible to AIMD that were previously thought not feasible. For that reason, developing such methods is an important objective and would have a major impact in scientific areas such as nanotechnology or biophysics, just to name a few. 

Several so called linear-scaling methods have been proposed \cite{goed, bowler1, baro, yang, galli, MGC, LNV, ManolopoulosPalser} to circumvent the cubic scaling diagonalization that is the main bottleneck of DFT and TB. Underlying all of these methods is the concept of "nearsightedness" \cite{kohn2, prod}, an intrinsic system dependent property, which states that at fixed chemical potential the electronic density depends just locally on the external potential, so that all matrices required to compute the Fermi operator will become sparse at last.
Together with sparse matrix algebra techniques linear scaling in terms of memory requirement and computational cost can be eventually achieved. 
However, the crossover point after which linear scaling methods become advantageous is still rather large, in particular for metallic systems and/or if high accuracy is needed. 

Therefore another method, based on the grand-canonical potential (GCP) for independent fermions, has been recently developed \cite{frenkel, alavi}. Krajewski and Parrinello demonstrated that by decomposing the GCP it is possible to devise an approximate stochastic linear scaling scheme \cite{kraj1, kraj2, kraj3}. Since this approach does not rely on the ability to localize the electronic wavefunction, even metals can be treated. However, due to its stochastic nature extending such a method towards self-consistent TB, DFT or HF is far from straightforward. 

This is where we start in this paper. Following previous work of Ceriotti, K\"uhne and Parrinello \cite{ceri1, ceri2} we compute here the finite-temperature density matrix, or Fermi matrix, in an efficient, accurate and in particular deterministic fashion by a hybrid approach. Inspired by the Fermi operator expansion method pioneered by Goedecker and coworkers \cite{goed, goed1, goed2}, the Fermi operator is described in terms of a Chebyshev polynomial expansion, but in addition is accompanied by fast summation as well as iterative matrix inversion techniques. 
The resulting algorithm has several important advantages. As before the presented scheme does rely on the ability to localize the orbitals, but requires only that the Hamiltonian matrix is sparse, a substantially weaker requirement. As a consequence not only metals, but even systems for which the Fermi matrix is not sparse yet can be treated with a linear scaling computational effort. Another advantage is that the algorithm is intrinsically parallel as the terms resulting from the decomposition of the GCP are independent of each other and can be separately calculated on different processors. 

But, at variance to the original approach \cite{kraj1, kraj2, kraj3}, the addition of Chebychev polynomial expansion and fast summation techniques leads to a particularly efficient algorithm that obeys a sub-linear scaling with respect to the width of the Hamiltonian's spectrum, which is very attractive for all-electron calculations or when a high energy resolution is required. Since the present method allows for an essentially exact decomposition of the GCP, without invoking any high-temperature approximation, it facilitates highly accurate linear scaling \textit{ab-initio} simulations. However, the main advantage lies in the deterministic nature of the hybrid approach, which enables self-consistent electronic structure calculations. 
The fact that the present scheme is based on the GCP inherently entails finite electron temperature, which is not only in line with finite temperature simulations such as AIMD, but furthermore also allows for computations of systems with excited electrons \cite{silvestrelli1, silvestrelli2}. 
In the present work, we have thus put particular emphasis on adopting the hybrid approach within AIMD. Specifically, the modified Car-Parrinello-like propagation of the self-consistent Hamilton matrix \cite{kueh1} and how to accurately sample the Boltzmann distribution with noisy forces \cite{kraj2, kueh1} are discussed in detail. 
Beside describing the method itself, we will show that it is indeed possible to perform fully self-consistent AIMD simulations and demonstrate the present scheme on liquid methane at planetary pressure and temperature conditions.

This article is organized as follows. In section II we summarize the basic methodology, first proposed by Krajewski and Parrinello \cite{kraj1, kraj2, kraj3}, while in section III we circumstantiate the novel hybrid approach. In Section IV the implementation within a self-consistent AIMD framework is described, while section V is devoted to the application on liquid methane at high temperature and pressure, as well as to the analysis of the actual computational complexity with respect to system size.

\section{\label{sec:level2}Basic Methodology}

We begin with the generic expression for the total energy $E$ of an effective single-particle theory, such as HF, DFT or TB 
\begin{equation}
  E = 2 \sum_{i=1}^{N} \varepsilon_{i} + V_{dc}. \label{TotEner}
\end{equation}
The first term denotes the so-called band-structure energy, which is given by the sum of the lowest $N$ doubly occupied eigenvalues $\varepsilon_{i}$ of an arbitrary Hamiltonian $\bm{H}$. In DFT for instance, $\bm{H}$ is the Kohn-Sham matrix, while $V_{dc}$ accounts for double counting terms as well as for the nuclear Coulomb interaction. In TB and other semi-empiricial theories $\bm{H}$ depends parametrically only on the nuclear positions and $V_{dc}$ is a pairwise additive repulsion energy. While in either case it is well known how to calculate $V_{dc}$ with linear scaling computational effort, the computation of all occupied orbitals by diagonalization requires $\mathcal{O}(N^{3})$ operations. Due to the fact that the band-structure term can be equivalently expressed in terms of the density matrix $\bm{P}$, the total energy can be written as 
\begin{equation}
  E = 2 \sum_{i=1}^{N} \varepsilon_{i} + V_{dc} = \text{Tr} [ \bm{P} \bm{H} ] + V_{dc}. \label{TotEnerMat}
\end{equation}
As a consequence, the cubic scaling diagonalization of $\bm{H}$ can be bypassed by directly calculating $\bm{P}$ rather than all $\varepsilon_{i}$'s. 

To that extend, we follow Alavi and coworkers \cite{frenkel, alavi} and consider the following (Helmholtz) free energy functional 
\begin{equation}
  \mathcal{F} = \Omega + \mu N_{e} + V_{dc}, \label{FreeEnerFunc}
\end{equation}
where $\mu$ is the chemical potential, $N_{e} = 2N$ the number of electrons and $\Omega$ the GCP for noninteracting fermions
\begin{eqnarray}
  \Omega &=& -\frac{2}{\beta}\, \textup{ln~det} \left( \bm{1} + e^{\beta\left(\mu \bm{S} - \bm{H} \right)} \right) \nonumber \\ 
  &=& - \frac{2}{\beta}\, \text{Tr}~\textup{ln} \left( \bm{1} + e^{\beta\left(\mu \bm{S} - \bm{H} \right)} \right). \label{GCPbasic}
\end{eqnarray}
Here, $\bm{S}$ stands for the overlap matrix, which is equivalent to the identity matrix $\bm{I}$ if and only if the orbitals are expanded in mutually orthonormal basis functions.
In the GCP the electronic temperature is finite and given by $\beta^{-1} = k_{B}T_{e}$. However, in the low-temperature limit 
\begin{equation}
  \lim_{\beta \rightarrow \infty} {\Omega} = 2 \sum_{i=1}^{N} {\varepsilon_{i}} - \mu N_{e} \label{GCPlowTempLim}
\end{equation}
the band-structure energy can be recovered and $\lim_{\beta \rightarrow \infty} {\mathcal{F}} = E$ holds. 
In order to make further progress, let us now factorize the operator of Eq.~(\ref{GCPbasic}) into $P$ terms. Given that $P$ is even, which we shall assume in the following, Krajewski and Parrinello \cite{kraj2, kraj3} derived the following identity 
\begin{eqnarray}
\bm{1} + e^{\beta\left(\mu \bm{S} - \bm{H} \right)} &=& \prod_{l=1}^{P} \left(\bm{1} - e^{\frac{i\pi}{2P}\left(2l-1\right)} e^{\frac{\beta}{2P}\left(\mu \bm{S} - \bm{H} \right)}\right) \nonumber \\
&=& \prod_{l=1}^{P} \bm{M}_l = \prod_{l=1}^{P/2} \bm{M}_l^*\bm{M}_l, \label{GCPdecomp}
\end{eqnarray}
where the matrices $\bm{M}_l$, with $l = 1, \ldots, P$ are defined by
\begin{equation} 
  \bm{M}_l := \bm{1} - e^{\frac{i\pi}{2P}\left(2l-1\right)} e^{\frac{\beta}{2P}\left(\mu \bm{S} - \bm{H} \right)}, \label{MlMat}
\end{equation}
while $^*$ denotes complex conjugation. Similar to numerical path-integral calculations, it is possible to exploit the fact that if $P$ is large enough, so that the effective temperature $\beta/P$ is small, the exponential operator $e^{\frac{\beta}{2P}\left(\mu \bm{S} - \bm{H} \right)}$ can be approximated by a Trotter decomposition or simply by a high-temperature expansion, i.e. 
\begin{equation}
  \bm{M}_{l} = \bm{1}-e^{\frac{i\pi}{2P}\left(2l-1\right)} \left( \bm{1} + \frac{\beta}{2P}(\mu \bm{S} - \bm{H}) \right) + \mathcal{O} \left( \frac{1}{P^2} \right). \label{approxMlMat}
\end{equation}
However, as we will see, here no such approximation is required, which is in contrast to the original approach \cite{kraj1, kraj2, kraj3}. 
In any case, the GCP can be rewritten as 
\begin{eqnarray}
\Omega &=& -\frac{2}{\beta} \textup{ln~det} \prod_{l=1}^{P} \bm{M}_l = -\frac{2}{\beta} \textup{ln} \prod_{l=1}^{P/2} \textup{det}\, (\bm{M}_l^*\bm{M}_l) \nonumber \\
&=& -\frac{2}{\beta} \sum_{l=1}^{P/2} \textup{ln~det}\, (\bm{M}_l^*\bm{M}_l) \nonumber \\
&=& \frac{4}{\beta} \sum_{l=1}^{P/2} \textup{ln} \left( \textup{det}\, (\bm{M}_l^*\bm{M}_l) \right)^{-\frac{1}{2}}. \label{GCP}
\end{eqnarray}

As is customary in lattice gauge field theory [\onlinecite[p. 17]{mont}], where the minus sign problem is avoided by sampling a positive definite distribution, the inverse square root of the determinant can be written as an integral over a complex field $\bm{\phi}_{l}$, which has the same dimension $M$ as the full Hilbert space, i.e. 
\begin{eqnarray}
  \det \left( \bm{M}_{l}^{*} \bm{M}_{l} \right)^{-1/2} = \frac{1}{(2\pi)^{\frac{M}{2}}} \int d\phi_l \, e^{-\frac{1}{2} \phi_l^*\bm{M}_l^*\bm{M}_l \phi_l}. \label{InvDet}
\end{eqnarray}
Inserting Eq.~(\ref{InvDet}) into Eq.~(\ref{GCP}) we end up with the following field-theoretic expression for the GCP: 
\begin{eqnarray}
  \Omega &=& \frac{4}{\beta} \sum_{l=1}^{P/2} \textup{ln}\, \left[ \frac{1}{(2\pi)^{\frac{M}{2}}} \int d\phi_l \, e^{-\frac{1}{2} \phi_l^*\bm{M}_l^*\bm{M}_l \phi_l} \right] \nonumber \\
  &=& \frac{4}{\beta} \sum_{l=1}^{P/2} \textup{ln}\, \int d\phi_l \, e^{-\frac{1}{2} \phi_l^*\bm{M}_l^*\bm{M}_l \phi_l} + const., \label{GCPft}
\end{eqnarray}
where, as already mentioned, $M$ is the dimension of $\bm{M}_l^*\bm{M}_l$ and $\phi_l$ are appropriate vectors.

All physical relevant observables can be computed as functional derivatives of the GCP with respect to an appropriately chosen external parameter. For example, $N_{e} = - \partial \Omega / \partial \mu$ and $\lim_{\beta \rightarrow \infty}{\Omega} + \mu N_{e} = 2 \sum_{i=1}^{N} \varepsilon_{i}$, so that 
\begin{equation}
  E = \lim_{\beta \rightarrow \infty}{\mathcal{F}} =  2 \sum_{i=1}^{N} \varepsilon_{i} + V_{dc} = \frac{\partial (\beta \Omega)}{\partial \beta} - \mu \frac{\partial \Omega}{\partial \mu} + V_{dc}. \label{GCPtotEner}
\end{equation}
Since the functional derivative of the constant in Eq.~(\ref{GCPft}) is identical to zero, all physical interesting quantities can be computed analog to 
\begin{subequations}
\begin{widetext}
\begin{eqnarray}
\frac{\partial \Omega}{\partial \lambda} &=& \frac{4}{\beta} \sum_{l=1}^{P/2} \frac{\int d\phi_l \, -\frac{1}{2} \phi_l^* \left( \frac{\partial (\bm{M}_l^*\bm{M}_l)}{\partial \lambda} \right) \phi_l  e^{- \frac{1}{2} \phi_l^* \bm{M}_l^* \bm{M}_l \phi_l }}{\int d\phi_l \, e^{ -\frac{1}{2} \phi_l^* \bm{M}_l^* \bm{M}_l \phi_l}} \\
&=& -\frac{2}{\beta} \sum_{l=1}^{P/2}  \frac{\int d\phi_l \, \sum\limits_{i,j = 1}^d (\phi_l)_i^* \left( \frac{\partial (\bm{M}_l^*\bm{M}_l)}{\partial \lambda} \right)_{ij} (\phi_l)_j \, e^{ - \frac{1}{2} \phi_l^* \bm{M}_l^* \bm{M}_l \phi_l }}{\int d\phi_l \, e^{ -\frac{1}{2} \phi_l^* \bm{M}_l^* \bm{M}_l \phi_l}} \\
&=& -\frac{2}{\beta} \sum_{l=1}^{P/2} \sum_{i,j = 1}^d \left( \frac{\partial (\bm{M}_l^*\bm{M}_l)}{\partial \lambda} \right)_{ij} \frac{\int d\phi_l \, (\phi_l)_i^*(\phi_l)_j \, e^{ - \frac{1}{2} \phi_l^* \bm{M}_l^* \bm{M}_l \phi_l }}{\int d\phi_l \, e^{ -\frac{1}{2} \phi_l^* \bm{M}_l^* \bm{M}_l \phi_l}} \\
&=& -\frac{2}{\beta} \sum_{l=1}^{P/2}  \sum_{i,j = 1}^d \left( \frac{\partial (\bm{M}_l^*\bm{M}_l)}{\partial \lambda} \right)_{ij} (\bm{M}_l^* \bm{M}_l)^{-1}_{ij} \\
&=& -\frac{2}{\beta} \sum_{l=1}^{P/2} \text{Tr} \left[ (\bm{M}_l^* \bm{M}_l)^{-1} \frac{\partial (\bm{M}_l^*\bm{M}_l)}{\partial \lambda} \right] \label{GCPfuncDeriv} \\
&=&  -\frac{2}{\beta} \sum_{l=1}^{P} \text{Tr} \left[ \bm{M}_l^{-1} \frac{\partial \bm{M}_l}{\partial \lambda} \right]. \label{GCPfuncDerivSym}
\end{eqnarray}
\end{widetext}
\end{subequations}
Thereby, Eq.~(\ref{GCPfuncDeriv}) holds because of Montvay and M\"unster [\onlinecite[p. 18]{mont}], while Eq.~(\ref{GCPfuncDerivSym}) is due to the fact that beside being positive definite $\bm{M}_l^* \bm{M}_l$ is also symmetric. 

Unlike Eq.~(\ref{GCP}), the determination of $\Omega = \partial(\beta \Omega) / \partial \beta$ does no longer require to calculate the inverse square root of a determinant, but only the inverse of $\bm{M}_{l}$. But, since the inversion usually has to be performed $P$ times, the computational scaling has presumably a rather large prefactor. Nevertheless, as we will see later this can be much ameliorated and all but very few matrix inversions can be avoided. 
On the other hand, $\bm{M}_{l}$ is not only very sparse, since it obeys the same sparsity pattern as $\bm{H}$, but is furthermore also always better conditioned as the latter, so that all $\bm{M}_{l}^{-1}$ matrices are substantially sparser than the finite temperature density matix and thus can be efficiently determined \cite{ceri1, ceri2}.
Solving the $N_{e}$ sets of linear equations $\bm{M}_{l} \bm{\Phi}_{j}^{l} = \bm{\psi}_{j}$, where $\{ \bm{\psi}_{j} \}$ is a complete set of basis functions, the inverse can be exactly computed as $\bm{M}_{l}^{-1} = \sum_{j=1}^{N_{e}}{\bm{\phi}_{j}^{l} \bm{\psi}_{j}^{l}}$ within $\mathcal{O}(N^{2})$ operations. 

Comparing Eq.~(\ref{TotEnerMat}) with Eq.~(\ref{GCPlowTempLim}) it is easy to see that the GCP and similarly all physical significant observables can be written as the trace of a matrix product consisting of the Fermi matrix $\bm{\rho}$, which in the low-temperature limit is equivalent to $\bm{P}$. 
Specifically, $\Omega = \partial (\beta \Omega) / \partial \beta = \text{Tr}[\bm{\rho} \bm{H}] - \mu N_{e}$, but because at the same time $N_{e} = \text{Tr}[\bm{\rho} \bm{S}]$ holds, the former can be simplified to 
\begin{equation}
  \Omega = \text{Tr} [\bm{\rho} (\bm{H} - \mu \bm{S})], 
\end{equation}
where 
$\bm{S}= - \partial \bm{H} / \partial \mu$ and $\bm{\rho} = {\partial \Omega} / {\partial \bm{H}}$. 
As a consequence, the GCP and all its functional derivatives can be reduced to evaluate $\bm{\rho}$ based on Eq.~(\ref{GCPfuncDeriv}) with $\lambda = H_{ij}$. 
Using the identity 
\begin{eqnarray}
  \frac{\partial \bm{M}_{l}}{\partial H_{ij}} &=& - \frac{1}{2P} \left\{ (\bm{M}_{l} - \bm{1}) \beta + \beta (\bm{M}_{l} - \bm{1}) \right\}, 
\end{eqnarray}
for this particular case Eq.~(\ref{GCPfuncDeriv}) eventually equals to 
\begin{eqnarray}
\bm{\rho} = \frac{\partial \Omega}{\partial \bm{H}} &=& \frac{4}{P} \sum_{l=1}^{P/2} \left(\bm{1} - \bigl(\bm{M}_l^* \bm{M}_l \bigr)^{-1}\right) \nonumber \\
&=& \frac{2}{P} \sum_{l=1}^{P} \left(\bm{1} - \bm{M}_l^{-1}\right). \label{DensityMatrix}
\end{eqnarray}
In other words, the origin of the method is the notion that the density matrix, the square of the wavefunction at low temperature and the Maxwell-Boltzmann distribution at high temperature, can be decomposed into a sum of $\bm{M}_l^{-1}$ matrices, each at higher effective temperature $\beta/P$ and hence always sparse than $\bm{\rho}$. 
Yet, contrary to the original approach \cite{kraj1, kraj2, kraj3}, neither a Trotter decomposition nor a high-temperature expansion for Eq.~(\ref{MlMat}) has been used, so far everything is exact for any $P$. Nevertheless, beside the aforementioned reduction from cubic to quadratic scaling no computational savings have been gained either. Quite the contrary, at first sight it might even appear that this scheme, which requires to invert $P$ matrices, is less efficient than explicitly diagonalizing $\bm{H}$. However, as already mentioned, in the next section we are going to demonstrate that this can be circumvented for the most part by expressing all but very few matrix inversions through a Chebychev polynomial expansion. 

\section{\label{sec:level3}The Hybrid Approach}

In order to make further progress and to achieve an even more favorable scaling, one can either approximate the propagator $e^{\frac{\beta}{2P}\left(\mu \bm{S} - H\right)}$ of Eq.~(\ref{MlMat}), or exploit the fact that by increasing $P$ in Eq.~(\ref{DensityMatrix}) the matrix exponential and hence $\bm{M}_l^{-1}$ can be ever simpler exactly calculated. Specifically, we employed the squaring and scaling technique to compute matrix exponentials, i.e. $e^{A} = (e^{A/m})^{m}$ \cite{MatExp2003}, where we exploit the fact that $e^{A/m}$ is trivial to compute whenever $P$ is large. 
In an analysis of the $\bm{M}_l$ matrices we found that every $\bm{M}_l$ matrix is throughout  better conditioned than $\bm{H}$ \cite{ceri1}. For this it follows, that for all $l$, $\bm{M}_l^{-1}$ always exhibits less nonzero entries and is therefore much easier to compute than the inverse of $\bm{H}$, which would correspond to the complexity of calculating $\bm{\rho}$ directly. 

In addition, the method can be even more improved by recognizing that $\bm{H}$ is real as well as symmetric and that the equality 
\begin{equation}
  \bm{M}_l = \bm{M}_{P-l+1}^* \label{MlConj} 
\end{equation} 
holds. Therewith, only the real parts of the $\bm{M}_l$ matrices are required to compute $\bm{\rho}$ (see Appendix), which entails substantial savings in terms of computational cost and memory requirement. From this it follows that Eq.~(\ref{DensityMatrix}) can be further simplified to 
\begin{equation} 
  \bm{\rho} = \frac{2}{P} \sum_{l=1}^{P/2} \left(\bm{1} - \operatorname{Re}\bm{M}_l^{-1} \right), \label{NewDensityMatrix}
\end{equation}
where the upper limit of index $l$ is henceforth restricted to $P/2$. Moreover, it has been observed that just a handful of $\bm{M}_l$ matrices, where $l$ is close to $(P+1)/2$, are ill-conditioned and only for them the inversion is computationally cumbersome. All other $\bm{M}_l$ matrices having a smaller index are rather well-conditioned, so that the matrix inversion can be very efficiently performed by a Chebyshev polynomial expansion \cite{ceri1}. This is to say that $\bm{\rho}$ can always be written as a sum of $\bm{M}_l^{-1}$ matrices, which are throughout pretty much sparser than $\bm{\rho}$ itself. The latter is in fact true even if $\bm{\rho}$ is rather full, so that metalic systems can be very efficiently treated. 

These complementary properties of the $\bm{M}_l$ matrices immediately suggest the following hybrid approach. Thereby an optimal $\bar l$ is chosen such that $1 < \bar l < P/2$, where all $\bm{M}_l$ matrices with $l < \bar l$ are inverted by a Chebychev polynomial expansion and only otherwise for $l \geq \bar l$ by an iterative Newton-Schulz matrix inversion. As long as $\bm{M}_l$ is not ill-conditioned, the former has the advantage of being essentially independent of $P$, so that increasing $P$ will not increase the computational cost. Together with the fact that the number of ill-conditioned $\bm{M}_l$ matrices does only depend on the particular system and $\beta$, but again not on $P$, the present hybrid approach allows to employ an arbitrary large $P$ at basically no additional computational cost. In this way the decomposition of the GCP in Eq.~(\ref{GCP}) can be made exact in any order essentially for free. From this it follows that the electronic temperature $\beta^{-1}$ can be chosen to be rather low and is typically identical with the nuclear temperature. 

Furthermore, it is possible to rewrite $\operatorname{Re}\bm{M}_l^{-1}$ in the following way: 
\begin{equation} 
  \operatorname{Re} \bm{M}_l^{-1} = \frac{1}{2} \left( \bm{1} +\bigl(e^{\frac{\beta}{P}\left(\bm{H} - \mu \bm{S} \right)} - \bm{1} \bigr) \bm{N}_l^{-1} \right), \label{ReMlinv}
\end{equation} 
where $\bm{N}_l$ is a real valued matrix as defined in the Appendix. That is to say, that the whole problem can be reduced to invert $\bm{N}_{l}$. Pretty much as for the $\bm{M}_{l}$ matrix, if $\bm{N}_l$ is well-conditioned, its inverse can be expressed by a Chebyshev expansion. For this purpose let us rewrite $\bm{N}_l$ in terms of a shifted and scaled auxiliary matrix 
\begin{equation}
  \bm{X} = \frac{e^{\frac{\beta}{2P}\left(\bm{H} - \mu \bm{S} \right)} - z_{0}}{\zeta}, \label{auxX}
\end{equation}
whose spectrum lies between $-1$ and $1$. The corresponding shifting and scaling parameters $z_{0} = \left( e^{\, \varepsilon_{\text{max}} / 2P} + e^{\, \varepsilon_{\text{min}} / 2P} \right) / 2$ and $\zeta = \left( e^{\, \varepsilon_{\text{max}} / 2P} - e^{\, \varepsilon_{\text{min}} / 2P} \right) / 2$ are expressed in terms of the maximum and minimum eigenvalues of $\bm{H}$, i.e. by $\varepsilon_{\text{max}}$ and $\varepsilon_{\text{min}}$ [\onlinecite{ceri2}]. Since a rather crude estimate for $\varepsilon_{\text{max}}$ and $\varepsilon_{\text{min}}$ is sufficient, they can be efficiently approximated using Gershgorin's circle theorem \cite{gersh} as 
\begin{subequations}
\begin{eqnarray}
  \varepsilon_{\text{max}} &\ge& \max_{i} \left( H_{ii} + \sum_{i \ne j} {\| H_{ij} \|} \right) \label{GershgorinA} \\
  \varepsilon_{\text{min}} &\le& \min_{i} \left( H_{ii} - \sum_{i \ne j} {\| H_{ij} \|} \right). \label{GershgorinB}
\end{eqnarray}
\end{subequations}

The difference $\Delta \varepsilon = \varepsilon_{\text{max}} - \varepsilon_{\text{min}}$ corresponds to the spectral width of $\bm{H}$ in unit of $k_{B}T$, which is also known as the HOMO-LUMO gap. 
The condition number $\kappa(\bm{N}_{l}) \approx 1 + \Delta \varepsilon^{2} \pi^{-2}(P-2l)^{-2}$ is somewhat higher than $\kappa(\bm{M}_{l}) \approx 1 + \Delta \varepsilon \pi^{-1}(P-2l)^{-1}$, but is more rapidly declining with decreasing $l$. 

Therewith, for $l < \bar l$, we can approximate $\bm{N}_l^{-1}$ as a sum of Chebychev polynomials of $\bm{X}$ by 
\begin{equation} 
  \bm{N}_l^{-1} \approx \sum_{i=0}^{m_{C}(l)} c_{li} T_i(\bm{X}), \label{Nlinv}
\end{equation}
where $T_i$ are Chebyshev polynomials and $c_{li}$ the corresponding coefficients. The upper bound $m_{C}(l)$ and thus the number of terms in the summation to achieve a relative accuracy of $10^{-D}$ on $\bm{N}_{l}^{-1}$ is approximately 

\begin{equation} 
  m_{C}(l) \approx \frac{1}{2} + \frac{\Delta \varepsilon D \ln 10}{\pi (P - 2l)}. \label{ChebCost}
\end{equation} 

After having computed the inverse of all the well-conditioned $\bm{N}_l$ matrices, we have to deal with the very few ill-conditioned ones. As already indicated this is accomplished by the following Newton-Schulz iteration
\begin{equation} 
  \bm{A}_{k+1} = 2\bm{A}_k - \bm{A}_k \bm{N}_l \bm{A}_k, \quad k = 0,1, \ldots, \label{NewtonSchultz}
\end{equation} 
which converges quadratically to $\bm{N}_l^{-1}$ given that $\bm{A}_0$ is within the respective area of convergence \cite{Schultz33}. Even though for 
\begin{equation} 
  \bm{A}_0 = \bm{N}_l^*\bigl(\lVert \bm{N}_l \rVert_1 \lVert \bm{N}_l \rVert_{\infty} \bigr)^{-1}, \label{InitialGuess}
\end{equation} 
Eq.~(\ref{NewtonSchultz}) is already guaranteed to converge \cite{pan}, but the computation of $\bm{N}_{l}^{-1}$ becomes even more efficient with the availability of a good initial guess for the matrix inverse. Fortunately, we can make use of $\bm{N}_{\bar l+n-1}^{-1}$ as an initial guess for $\bm{N}_{\bar l+n}^{-1},~n \in \{0,\ldots, P/2 - \bar l \}$, that is good enough to even converge rather ill-conditioned matrices usually within a few iterations. The number of matrix multiplications required to obtain a relative accuracy of $10^{-D}$ on $\bm{N}_{l}$ starting from $\bm{N}_{\bar l-1}$ that has already been calculated by Eq.~(\ref{Nlinv}) is
  \begin{eqnarray} 
    m_{N}(l) &=& \frac{2}{\ln 2} \ln \frac{\ln (1-\chi(l)) - D \ln 10}{\ln \chi(l)}, ~\text{where} \nonumber \\ 
    \chi(l) &\approx& \frac{4(P+1-2l)}{\left(1+(P+1+2l)\right)^{2}}. \label{NewtonCost}
  \end{eqnarray} 

Hence, the optimal value of $\bar l$ can be found by minimizing the estimated total number of matrix multiplications 
\begin{equation} 
  m_{tot}(\bar l) = m_{C}(\bar l) + \sum_{l = \bar l}^{P/2}{m_{N}(l)} \label{TotCost}
\end{equation} 
under variation of $\bar l$. In general $P/2 - \bar l$ is rather small and only weakly dependent on $\beta$, which implies that just a few $\bm{N}_{l}$ matrices needs to be explicitly inverted using Eq.~(\ref{NewtonSchultz}), regardless of the electronic temperature. 

However, the matrix-matrix multiplications of Eq.~(\ref{NewtonSchultz}) causes that $\lim_{k \rightarrow \infty} \bm{A}_{k+1}$ eventually becomes fairly occupied. For this reason in order to sustain linear scaling the intermediate matrices are truncated. Nevertheless, as already mentioned, the condition number of $\bm{N}_{l}$ is always lower than the one of $\bm{H}$ and typically even rather well-conditioned, so that $\bm{N}_{l}^{-1}$ is by definition substantially sparser than $\bm{\rho}$. From this it follows that the necessary truncation cutoff is relatively mild and the approximation therefore very small, so that highly accurate linear scaling electronic structure calculations are still possible. 

\section{\label{sec:level4}Performing Self-Consistent \\ Molecular Dynamics Simulations}

The chief advantage of this procedure is not only that it allows for accurate linear scaling calculations, but is furthermore also deterministic. Hence, at variance to the original approach \cite{kraj1, kraj2, kraj3}, where the corresponding matrices are inverted by an approximate stochastic method, it is now possible to perform calculations using Hamilton operators of fully self-consistent mean-field theories, such as HF, DFT and self-consistent TB (SCTB) \cite{sutton, elstner}. 

We have tested the method in the context of electronic structure based molecular dynamics (MD) using a SCTB model \cite{horsf} and implemented it in the CMPTool program package \cite{meloni,cmpt}. In the self-consistent field (SCF) optimization loop self-consistency is realized by imposing local charge neutrality, to account for charge transfer processes, as well as bond breaking and formation. This means that the number of electrons of every atom $\alpha$ has to be equal to the number of its valence electrons $q_{\alpha}^{0}$ within an adjustable tolerance, which we named $\Delta q_{\operatorname{max}}$. To that extend, during the SCF loop the diagonal elements of $\bm{H}$ are varied using a linear response function $\bm{\Theta}$ until local charge neutrality is achieved. Specifically, in each MD step first $\bm{H}$ is built up, whereas in every SCF iteration we calculate the shift-vector $\bm{\Delta}_H$ to the diagonal elements of $\bm{H}$. The latter are the so called on-site energies $\epsilon_{i} = \bm{H}_{ii}$, while the diagonal elements of $\bm{\rho} \bm{S}$ represents the occupancy of the corresponding orbital, hence $N_{e} = \text{Tr}[\bm{\rho S}]$. Summing over all orbitals centered on any particular atom $\alpha$, one obtains the associated on-site charge $q_{\alpha}$. Local charge neutrality is enforced by calculating $\bm{\Delta}_{H}^{k} = \bm{\Theta} (q_{\alpha}^{k} - q_{\alpha}^{0})$ for every SCF iteration $k$ and shifting the on-site energies using $\epsilon_{i}^{k+1} = \epsilon_{i} + \bm{\Delta}_{H}^{k}$. So adapted, $\bm{H}^{k}$ is diagonalized using the above formalism until $\max_{\alpha} |q_{\alpha} - q_{\alpha}^{0}| \le \Delta q_{\operatorname{max}}$. In that case, instead of being grand-canonical the simulation is performed at constant $N_{e}$. 

However, as already recognized by Kress et al. \cite{kress1} using the present SCTB model \cite{horsf}, the SCF cycle is  very slowly converging and the number of necessary iterations critically dependent on $\Delta q_{\operatorname{max}}$. Nevertheless, this can remedied by adapting the method of K\"uhne et al. \cite{kueh1} in such a way that instead of a fully coupled electron-ion MD only the modified predictor-corrector integrator is used to propagate $\bm{\Delta}_H$ in time. In the framework of DFT this scheme has shown to be particularly effective for a large variety of different systems \cite{caraAPL, *caraJPCM1a, 
*caraPRL, *caraJPCM2, *losInSb, *losInSbTe, camellone, *cucinotta, *luduena1, *chao}. 
Inspired by the original scheme of Kolafa \cite{kolafa1, kolafa2} here 
\begin{equation}
\bm{\Delta}_H(t_n)^{p} = \sum_{m=1}^K (-1)^{m+1}m\frac{\binom{2K}{K-m}}{\binom{2K-2}{K-1}} \bm{\Delta}_H(t_{n-m}) \label{ASPCpredictor}
\end{equation}
is used as a modified predictor, where $\bm{\Delta}_H(t_n)^{p}$ is an estimate for $\bm{\Delta}_H(t_n)$ of the next MD time step $t_n$ and is approximated using the weighted shifts of the $K$ previous time steps. 
As we will show in the Appendix the weights 
\begin{equation}
  w_m = (-1)^{m+1}m\frac{\binom{2K}{K-m}}{\binom{2K-2}{K-1}}
\end{equation}  
always add up to $1$, so that $K=1$ causes that $w_1$ is identical to $1$, i.e. $\bm{\Delta}_H(t_n)^p = \bm{\Delta}_H(t_{n-1})$. 

However, contrary to the second generation Car-Parrinello MD approach of K\"uhne et al. \cite{kueh1}, 
where in each MD step only a single preconditioned electronic gradient calculation is required as the corrector, here the predicted $\bm{\Delta}_H(t_n)^{p}$ is only used as an initial guess for the SCF cycle, which requires at least a single if not multiple diagonalizations. That is to say that instead of a genuine Car-Parrinello-like dynamics \cite{kueh1, cpmd}, a less efficient accelerated Born-Oppenheimer MD (BOMD) \cite{arias1, alfe, pulay, herbert, quickstep, niklasson} is performed. 

Nevertheless, in this way the convergence rate of the SCF cycle is much increased, while at the same time even allowing for a rather tight tolerance threshold. In fact, comparing with the employed convergence criterion of Kress et al. \cite{kress1}, here $\Delta q_{\operatorname{max}}$ can be chosen to be at least one to two orders of magnitude smaller without requiring numerous SCF iterations. 

Due to the fact that the present scheme is equivalent to diagonalizing $\bm{H}$, as for any SCF theory based BOMD simulation, the interatomic forces thus calculated are affected by a statistical noise $\bm{\Xi}_{I}^{N}$, except for the unrealistic case that $\Delta q_{\operatorname{max}} = 0$. 
Hence, instead of the exact forces $\bm{F}_{I}$, merely an approximation $\bm{F}_{I}^{\text{BOMD}} = \bm{F}_{I} + \bm{\Xi}_{I}^{N}$ is computed, where $\bm{F}_{I}^{\text{BOMD}}$ are the BOMD forces calculated by an arbitrary SCF based theory. 
Even though, $\bm{\Xi}_{I}^{N}$ can, to a very good approximation, be assumed as white \cite{kraj2,dai}, the line integral defining the net work is always positive and thus entails an energy drift during a microcanonical MD simulation. While the noise may be tiny and the forces highly accurate, as far as static calculations such as geometry optimization are concerned, the resulting energy drift is way more critical. An energy drift of as small as 1~$\mu$eV/(atom$\times$ps) grows to an aberration of 10~K/ns and may cause that liquid water for instance evaporates within a couple of nanoseconds simply because of the energy drift immanently present in any BOMD simulation \cite{kuehne1, *kuehne2, *pascal, *kuehne3, *WaterReview}. Therefore, at least in principle, it is no longer guaranteed that by solving Newton's equation of motion (EOM) the correct Boltzmann averages are obtained.

Fortunately, only based on the assumption that $\bm{\Xi}_{I}^{N}$ is unbiased, this can be rigorously corrected by devising a modified Langevin equation \cite{kraj2, kueh1}. 
Specifically, taking cue from the work of Krajewski and Parrinello, we sample the canonical distribution using the following equation: 
\begin{subequations}
  \begin{eqnarray}
    M_{I} \ddot{\bm{R}}_{I} &=& \bm{F}_{I} + \bm{\Xi}_{I}^{N} - \gamma_{N} M_{I} \dot{\bm{R}}_{I} \label{ModLangevinEqA} \\
&=& \, \bm{F}_{I}^{\text{BOMD}} \, - \, \gamma_{N} M_{I} \dot{\bm{R}}_{I}, \label{ModLangevinEqB}
  \end{eqnarray}
\end{subequations}
where $M_{I}$ is the nuclear mass and $\gamma_{N}$ a friction coefficient to compensate for the noise $\bm{\Xi}_{I}^{N}$. 
The latter has to obey 
\begin{eqnarray}
  \langle \bm{F}_{I}(0) \bm{\Xi}_{I}^{N}(t) \rangle &\cong& 0, \label{whiteNoise}
\end{eqnarray}
as well as the so called fluctuation-dissipation theorem
\begin{eqnarray}
  \langle \bm{\Xi}_{I}^{N}(0) \bm{\Xi}_{I}^{N}(t) \rangle &=& 2\gamma_{N} k_{B}TM_{I} \delta(t). \label{FlucDissTheorem}
\end{eqnarray}
If we would know $\gamma_{N}$ such that Eq.~(\ref{FlucDissTheorem}) is satisfied, a genuine Langevin equation is recovered, which guarantees for an accurate canonical sampling of the Boltzmann distribution. 
However, at first sight this may look like an impossible undertaking, since we neither know $\bm{F}_{I}$, nor $\bm{\Xi}_{I}^{N}$ from which $\gamma_{N}$ can be deduced. Nevertheless, it is possible, even without knowing $\bm{\Xi}_{I}^{N}$ except that it is approximately unbiased, to determine $\gamma_{N}$ directly by simply varying it in such a way that the equipartition theorem $\left\langle \frac{1}{2} M_{I} \dot{\bm{R}}_{I}^{2} \right\rangle = \frac{3}{2} k_{B} T$ holds. Once $\gamma_{N}$ is determined, it must be kept constant for the whole simulation. But then it is possible to exactly and very efficiently calculate static and even dynamic observables without knowing $\bm{F}_{I}$, but just $\bm{F}_{I}^{\text{BOMD}}$. Due to the fact that the same also holds for the noise introduced by truncating the intermediate matrices of Eq.~(\ref{Nlinv}), as well as using finite-precision arithmetic and a non-vanishing integration time step, the corresponding noise terms can be simply added to $\bm{\Xi}_{I}^{N}$. 

\section{\label{sec:level5}Liquid methane at extreme pressure and temperature}

For the purpose of demonstrating the present method, we present here initial results on liquid CH$_4$ at high pressure and temperature. All BOMD simulations have been performed using the SCTB model for hydrocarbons \cite{horsf} as implemented in the CMPTool code \cite{cmpt}. The calculations have been performed in the canonical ensemble at $T = 2000$~K and volume $V=10.04$~cm$^3$/mol, which corresponds to the second-shock at pressure $P = 92$~GPa of a two-stage light-gun shock compression experiment \cite{nellis1}. The EOM of Eq.~(\ref{ModLangevinEqB}) is integrated using a discretized time step of $\Delta t = 0.5$~fs. 
We are considering here a periodic cubic box of length $L_1 = 12{.}775$~\r{A}, consisting of 125 liquid CH$_4$ molecules as our unit cell. The local charge neutrality threshold of the SCF loop $\Delta q_{\text{max}} = 0.05$. Since we are dealing with a large disordered system at finite temperature, the Brillouin zone is sampled at the $\Gamma$-point only. The electronic temperature is equal to the nuclear temperature, i. e. $T_{e}=2000~K$. The minimization of Eq.~(\ref{TotCost}) with respect to $\bar l$ yields $\bar l = P/2 - 2$, which implies that all except for two $\bm{N}_{l}$ matrices can be efficiently computed by a Chebychev polynomial expansion with an estimated $m_{C}(\bar l) \leq 61$. Nevertheless, since Eq.~(\ref{TotCost}) is merely an approximation, in practice the overall efficiency can be further increased by reducing $\bar l$. Here we have employed $\bar l = P/2-4$, which results in $m_{C}(\bar l) \approx 30$. 

\begin{figure}
\includegraphics[width=0.475\textwidth]{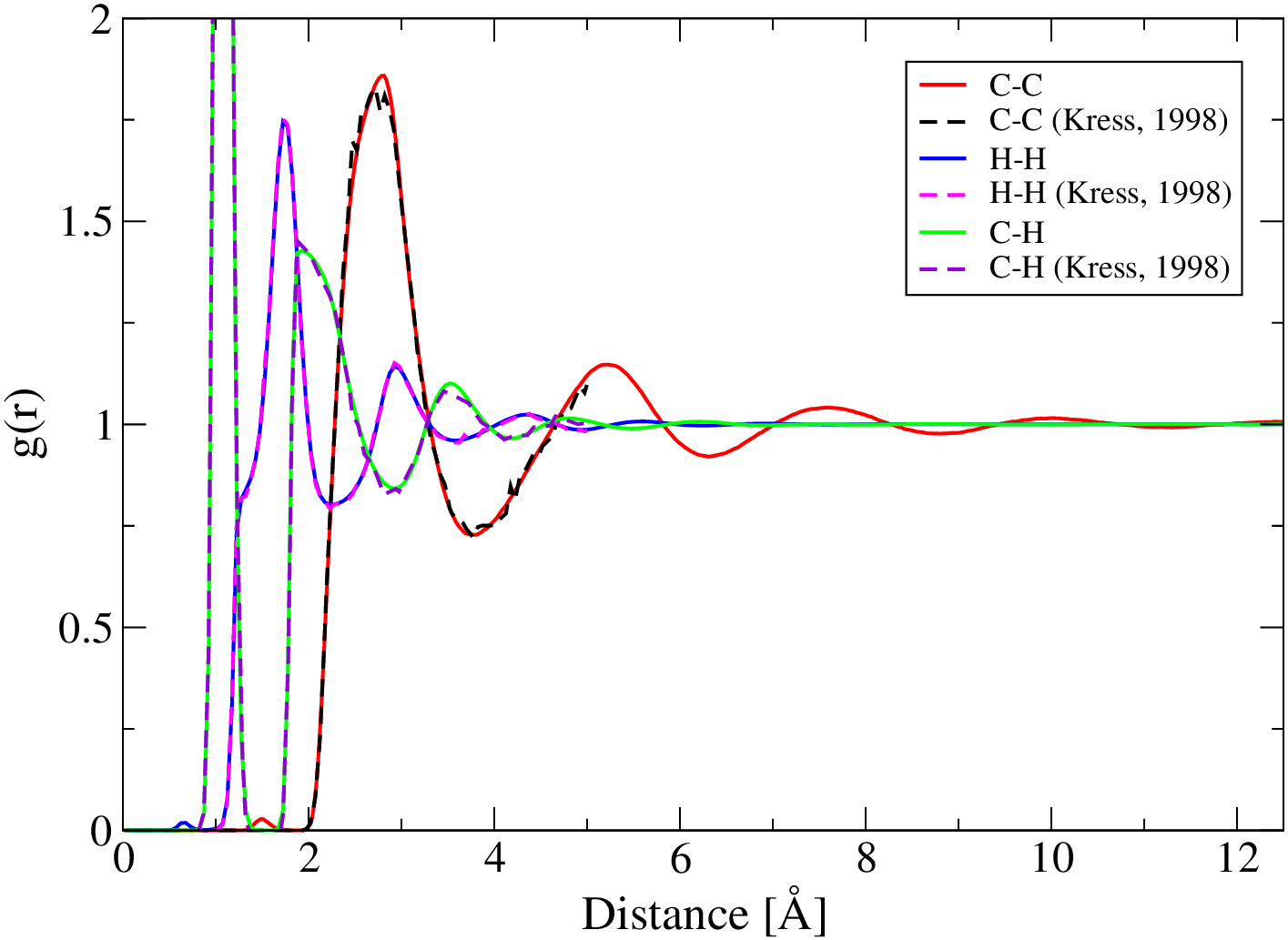}
\caption{(Color online) Partial pair-correlation functions g(r) of liquid CH$_4$ at 2000~K.}
\label{fig:pcf}
\end{figure}

\begin{figure}
\includegraphics[width=0.475\textwidth]{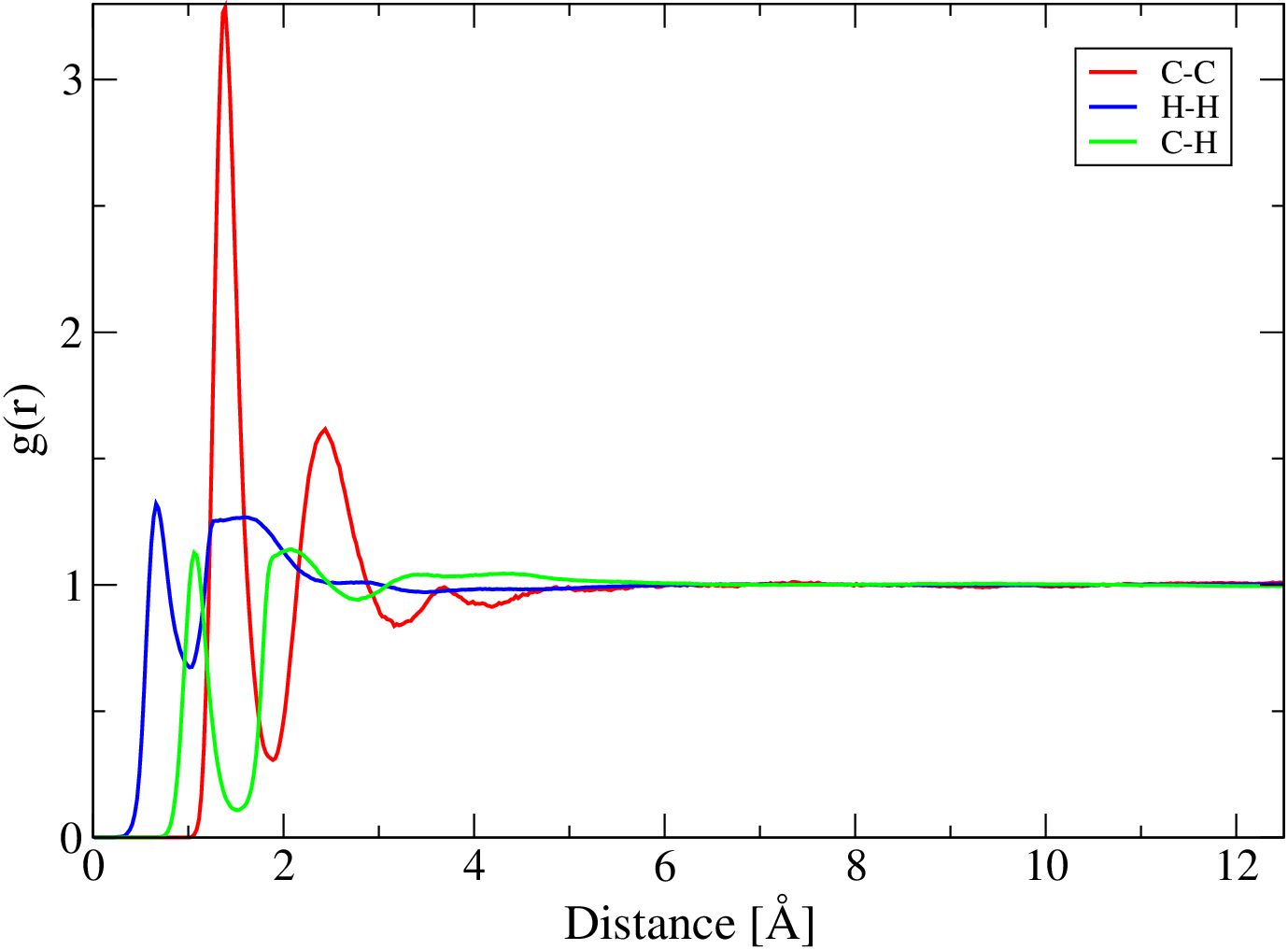}
\caption{(Color online) Partial pair-correlation functions g(r) of liquid CH$_4$ at 8000~K.}
\label{fig:pcf_8000}
\end{figure}

To assess the accuracy we study a sample comprising of 1000 CH$_4$ molecules (2$\times$2$\times$2 the size of our unit cell) at $T=2000$~K and compare the partial pair-correlation functions, 
as obtained by the present scheme, with the results of Kress et al. \cite{kress1} using exactly the same model \cite{horsf}. As can be seen in Fig.~\ref{fig:pcf} the agreement is excellent. 
The fact that even metallic systems can be treated with linear system size scaling is demonstrated on exactly the same system at $T=8000$~K. We find that at this temperature the CH$_4$ molecules are partially dissociated, as indicated by the reduced intramolecular C-H peak in Fig.~\ref{fig:pcf_8000}. Similar, from the first C-C and H-H peaks, the occurrence of covalent C-C bonds and H$_2$ molecules can be deduced. Moreover, a noticeable fraction of monoatomic hydrogen can be identified, which immediately suggests that hydrogen is on the verge of a liquid-liquid phase transition into an atomic fluid phase that is in agreement with recent AIMD calculations \cite{Tamblyn2010, Morales2010}. Eventually, the electronic band-gap is vanishing, which is most likely due to the emergence of monoatomic hydrogen. Further calculations to investigate the dissociation of methane and its implication for giant gas planets such as Uranus and Neptune will be discussed in a forthcoming paper \cite{richMethane}.

In order to sustain linear scaling in terms of computational cost and at the same time and memory requirement, all sparse matrices are stored in the common Compressed Row Storage format. 
Due to the fact that the algorithm heavily relies on the multiplication of sparse matrices, we have put particular emphasis on an efficient parallel implementation. In that the data is distributed to the individual processor cores by employing a space-filling Hilbert curve to keep the load balanced \cite{bowler2}. While for solid state systems a very good scalability has been observed, for disordered liquids studied here the situation is substantially less favorable. A more efficient scheme, which dynamically rearranges the matrices between the various processor cores, or even distributes them fully at random is current work in progress.

\begin{figure}
\includegraphics[width=0.475\textwidth]{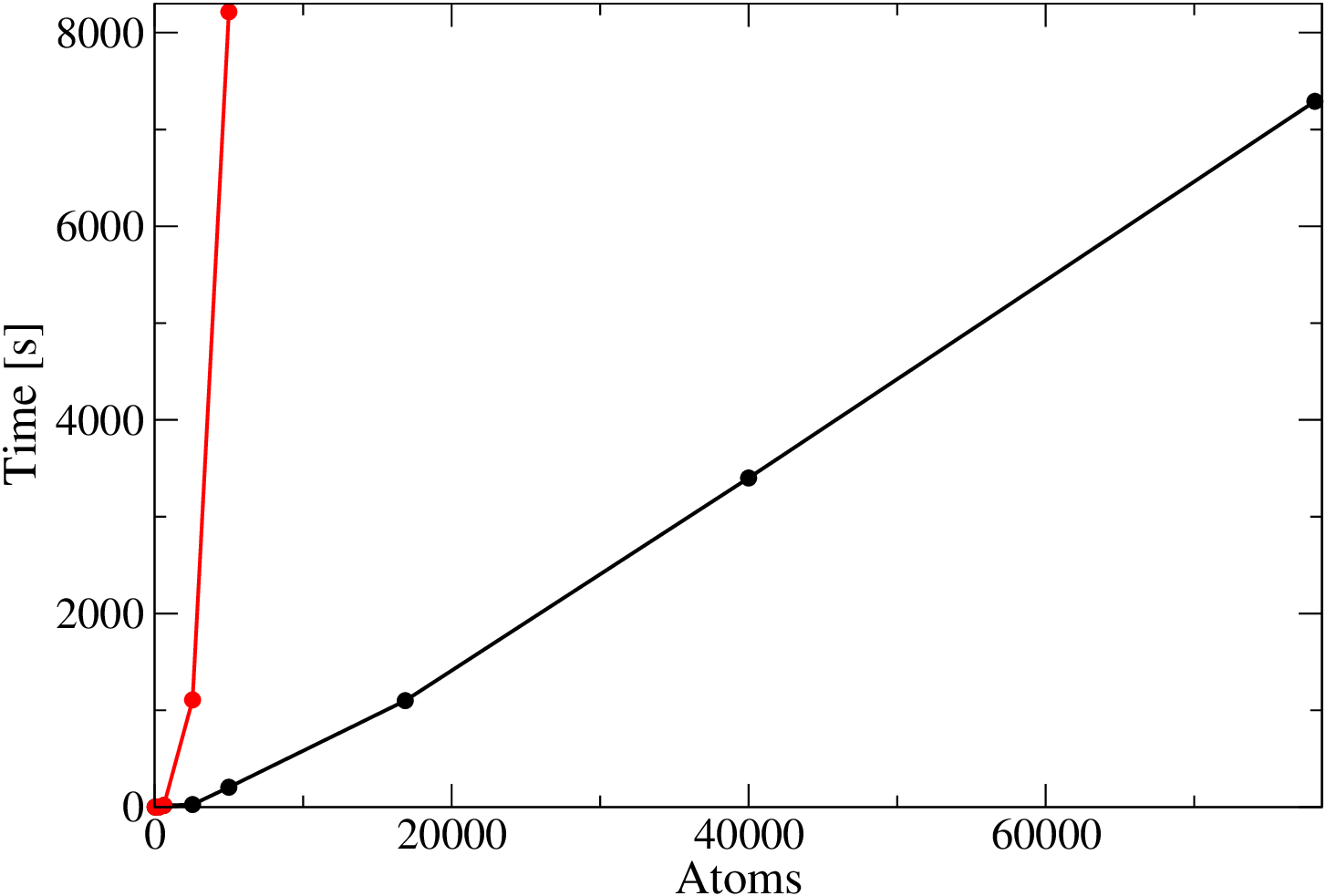}
\caption{(Color online) The average walltime for a single SCTB MD step versus the number of atoms on a single core of a 2.40~GHz Intel Westmere processor. The walltime using a divide and conquer diagonalization algorithm is shown in red, while the present linear scaling scheme is denoted in black.}
\label{fig:runtime}
\end{figure}

Nevertheless, to demonstrate that linear system size scaling is indeed attained, in Fig. \ref{fig:runtime} the average runtime for a complete SCTB MD step at $T = 8000$~K is shown for various system sizes using a single core of a 2.40~GHz Intel Westmere processor. Specifically, we have considered five different systems, beginning with our unit cell up to 5$\times$5$\times$5 replications of it, which corresponds to 625, 5000, 16875, 40000 and 78125 atoms, respectively. As can be seen in Fig. \ref{fig:runtime}, for small system sizes up to around 10000 atoms the scaling is even sub-linear and thereafter perfectly linear with system size. Comparing the runtime with a divide and conquer diagonalization algorithm unveils that the crossing point, after which the linear scaling algorithm becomes computationally more favorable, is at $N_{C} \approx 425$ atoms. 

However, beside the formal scaling with system size, the corresponding prefactor is also rather important and depends on $\Delta \varepsilon$. 
In the case of Chebychev polynomial based Fermi operator expansion methods, the computational cost has been found to scale like $D \beta \Delta \varepsilon$, in order to achieve an accuracy of $10^{-D}$ [\onlinecite{baer}]. Apparently, this entails a fairly large prefactor if either high accuracy is required, the electronic temperature is low, or when $\Delta \varepsilon$ is large. The latter is typically the case for an all-electron calculation, or if a plane wave basis set is employed. Nevertheless, the usage of fast polynomial summation methods leads to the more favorable scaling $\sqrt{\beta \Delta \varepsilon}$ [\onlinecite{VanLoan}, \onlinecite{liang1}, \onlinecite{liang2}]. For the present hybrid approach this results in an even better sub-linear scaling of $\sqrt[3]{\beta \Delta \varepsilon}$ [\onlinecite{ceri2}], which makes it particularly attractive for highly accurate all-electron \textit{ab-initio} calculations, or when a high energy resolution is required.  Together with the methods proposed by Lin et al., this is the best scaling with respect to $\beta$ and $\Delta \varepsilon$ reported so far \cite{lin1, lin2}. 
Based on a multipole representation of the Fermi operator, the latter scales as $\ln(\beta \Delta \varepsilon) \ln(\ln(\beta \Delta \varepsilon))$, which depending on the actual value of $\beta \Delta \varepsilon$ is either slightly lower or larger than the present cubic root scaling. 

\section{\label{sec:level6} Conclusion}

In conclusion, we would like to mention, that the here presented method can be directly applied to fully self-consistent DFT calculations 
by writing $V_{dc}$ of Eq.~(\ref{TotEner}) as 
\begin{eqnarray}
  V_{dc}[\bm{\rho}(\bm{r})] &=& - \frac{1}{2} \int{ \, d\bm{r} \int{ \, d\bm{r}' \frac{\bm{\rho}(\bm{r})  \bm{\rho}(\bm{r}') }{|\bm{r}-\bm{r}'|} } } \nonumber \\
  &-& \int{\, d\bm{r} \, \bm{\rho}(\bm{r}) \frac{\delta \Omega_{\text{XC}}}{\delta \bm{\rho}(\bm{r})}} + \Omega_{\text{XC}} + E_{II}, \label{dcDFT}
\end{eqnarray}
where the first term on the right hand side is the double counting correction of the Hartree energy, while $\Omega_{\text{XC}}$ is the finite-temperature exchange and correlation grand-canonical functional and $E_{II}$ the nuclear Coulomb interaction. Except for the latter term, Eq.~(\ref{dcDFT}) accounts for the difference between the GCP for independent fermions $\Omega$ and the GCP for the interacting spin-$\frac{1}{2}$ Fermi gas 
\begin{eqnarray}
  \Omega_{int}[\bm{\rho}(\bm{r})] &=& -\frac{2}{\beta} \ln \det \left( \bm{1} + e^{\beta\left(\mu \bm{S} - \bm{H} \right)} \right) \nonumber \\
  &-& \frac{1}{2} \int{ \, d\bm{r} \int{ \, d\bm{r}' \frac{\bm{\rho}(\bm{r})  \bm{\rho}(\bm{r}') }{|\bm{r}-\bm{r}'|} } } \nonumber \\
  &-& \int{\, d\bm{r} \, \bm{\rho}(\bm{r}) \frac{\delta \Omega_{\text{XC}}}{\delta \bm{\rho}(\bm{r})}} + \Omega_{\text{XC}}.
\end{eqnarray}
As before, in the low-temperature limit $\Omega_{int}[\bm{\rho}(\bm{r})] + \mu N_{e}$ equals to the band-structure energy, whereas $\Omega_{\text{XC}}$ corresponds to the familiar exchange and correlation energy, so that in this limit $\mathcal{F} = \Omega + \mu N_{e} + V_{dc} = \Omega_{int}[\bm{\rho}(\bm{r})] + \mu N_{e} + E_{II}$ is equivalent to the Harris-Foulkes energy functional \cite{harris, foulkes}.
Such as the latter, $\mathcal{F}$ is explicitly defined for any $\bm{\rho}(\bm{r})$ and obeys exactly the same stationary point as the finite-temperature functional of Mermin \cite{mermin}. 

The formal analogy of the decomposition to the Trotter factorization immediately suggests the possibility to apply some of the here presented ideas with benefit to numerical path-integral calculations \cite{ceperley}. 
The same applies for a related area where these methods are extensively used, namely the lattice gauge theory to quantum chromodynamics \cite{kogut}, whose action is rather similar to the one of Eq.~(\ref{InvDet}). 

\begin{acknowledgments}
The authors would like to thanks Michele Ceriotti, Martin Hanke-Bourgeois and Luca Ferraro for valuable suggestions regarding the implementation of the presented method. 
Financial support from the Graduate School of Excellence MAINZ, the Max Planck Graduate Center and the IDEE project of the Carl Zeiss Foundation is kindly acknowledged. 
\end{acknowledgments}

\section{Appendix}

\subsection{\label{app:A} Proof that $\bm{M}_l^* \bm{M}_l \in \mathbbm{R}^{D \times D}$}

Using Eq.~\eqref{MlConj} and the fact that $\omega_l := e^{\frac{i\pi}{2P}\left(2l-1\right)}$ denotes a point on the unit circle of the complex plane
\begin{align}
\bm{M}_l^* \bm{M}_l =& \left(\bm{1} - \omega_l e^{\frac{\beta}{2P}\left(\mu \bm{S} - \bm{H} \right)}\right)^*\left(\bm{1} - \omega_l e^{\frac{\beta}{2P}\left(\mu \bm{S} - \bm{H} \right)}\right) \nonumber \\
=& \left(\bm{1}^* - \overline{\omega}_l \bigl(e^{\frac{\beta}{2P}\left(\mu \bm{S} - \bm{H} \right)}\bigr)^*\right)\left(\bm{1} - \omega_l e^{\frac{\beta}{2P}\left(\mu \bm{S} - \bm{H} \right)}\right) \nonumber \\
=&~ \bm{1} - (\overline{\omega}_l + \omega_l) e^{\frac{\beta}{2P}\left(\mu \bm{S} - \bm{H} \right)} + \left(\overline{\omega}_l \omega_l \right) e^{\frac{\beta}{P}\left(\mu \bm{S} - \bm{H} \right)} \nonumber \\
=&~ \bm{1} - 2\operatorname{Re} \omega_l\, e^{\frac{\beta}{2P}\left(\mu \bm{S} - \bm{H} \right)}+e^{\frac{\beta}{P}\left(\mu \bm{S} - \bm{H} \right)} \nonumber \\
=& \left(\bm{1} + e^{\frac{\beta}{P}\left(\bm{H} - \mu \bm{S} \right)} - 2\operatorname{Re} \omega_l \, e^{\frac{\beta}{2P}\left(\bm{H} - \mu \bm{S} \right)}\right) e^{\frac{\beta}{P}\left(\mu \bm{S} - \bm{H} \right)} \nonumber \\
=:&~\bm{N}_l\, e^{\frac{\beta}{P}\left(\mu \bm{S} - \bm{H} \right)} \in \mathbbm{R}^{M \times M}
\end{align}
where $M$ is the number of basis functions and therefore the dimension of the real matrix $\bm{M}_l^* \bm{M}_l$. 

\subsection{\label{app:B} Proof that $\sum_{m=1}^K w_m = 1$}

For the purpose to show that
\begin{equation} 
  \sum_{m=1}^K w_m = \sum_{m=1}^K (-1)^{m+1}m\frac{\binom{2K}{K-m}}{\binom{2K-2}{K-1}} = 1,  \label{eqn:app} 
\end{equation} 
we make use of the Appendix of Ref.~(\onlinecite{kolafa1}) and write
\begin{widetext}
  \begin{subequations}
    \begin{align}
      &\sum_{m=1}^K (-1)^{m+1}m\binom{2K}{K-m} \\
      =& \sum_{m=1}^K (-1)^{m+1}m\left[ \binom{2K-2}{K-m} + 2 \binom{2K-2}{K-m-1} + \binom{2K-2}{K-m-2} \right] \\
      =& \sum_{m=1}^K (-1)^{m+1} \left[\frac{m(2K-2)!}{(K-m)!(K+m-2)!} + \frac{2m(2K-2)!}{(K-m-1)!(K+m-1)!} + \frac{m(2K-2)!}{(K-m-2)!(K+m)!}  \right]. 
    \end{align}
  \end{subequations}
\end{widetext}
All but the first summand cancels out so that we get
\begeq{\sum_{m=1}^K (-1)^{m+1}m\binom{2K}{K-m} = \binom{2(K-1)}{K-1}.}  
which after inserting it into Eq.~\eqref{eqn:app} equals to 
\begeq{\sum_{m=1}^K w_m = \binom{2K-2}{K-1}^{-1} \cdot \binom{2(K-1)}{K-1} = 1.}


%

\end{document}